\documentclass[aps,reprint,twocolumn,showpacs,preprintnumbers,amsmath,amssymb,prl,footinbib,10pt,floatfix]{revtex4-1}

\pdfoutput=1
\usepackage{pdfpages} 

\usepackage{graphicx}
\usepackage{dcolumn}
\usepackage[tight]{subfigure}
\usepackage{amsmath}
\usepackage{verbatim}
\usepackage{color}
\usepackage{bm} 
\usepackage{natbib}
\usepackage{xspace}
\usepackage{physics}
\usepackage{mathtools}
\usepackage{hyperref} \hypersetup{colorlinks=true,linktoc=all,linkcolor=blue,breaklinks=true,citecolor=blue,urlcolor=blue}
\usepackage{etoolbox}
\usepackage{epstopdf}

\makeatletter
\patchcmd{\@outputpage@head}{\@ifx{\LS@rot\@undefined}{}{\LS@rot}}{}{}{}
\makeatother

\begin{document}
\title{Non-Adiabatic Dynamics in Single-Electron Tunneling Devices with Time-Dependent Density Functional Theory}
\author{Niklas Dittmann$^{(1,2,3)}$}
\author{Janine Splettstoesser$^{(1)}$}
\author{Nicole Helbig$^{(3)}$}
\affiliation{
  (1) Department of Microtechnology and Nanoscience (MC2), Chalmers University of Technology, SE-41298 G{\"o}teborg, Sweden\\
  (2) Institute for Theory of Statistical Physics, RWTH Aachen, 52056 Aachen, Germany\\
  (3) Peter-Gr\"unberg Institut and Institute for Advanced Simulation, Forschungszentrum J\"ulich, 52425 J\"ulich, Germany\\
}


\begin{abstract}
We simulate the dynamics of a single-electron source, modeled as a quantum dot with on-site Coulomb interaction and tunnel coupling to an adjacent lead, in time-dependent density functional theory. 
Based on this system, we develop a time-nonlocal exchange-correlation potential by exploiting analogies with quantum-transport theory. The time non-locality manifests itself in a dynamical potential step. 
We explicitly link the time evolution of the dynamical step to \emph{physical} relaxation time scales of the electron dynamics.
Finally, we discuss prospects for simulations of larger mesoscopic systems.
\end{abstract}
\maketitle


Time-dependent density functional theory (TDDFT) is a widely-used tool for the calculation of time-dependent phenomena in interacting quantum systems \cite{MarquesMaitraNogueiraGrossRubio12,Ullrich11,Maitra16}.
In TDDFT, the time-dependent electronic density of interacting electrons moving in an external potential $v(\mathbf{r},t)$ is calculated
from non-interacting electrons, which are placed in the artificial Kohn-Sham (KS) potential $v_\mathrm{KS}(\mathbf{r},t)$.
The difference, $v_\mathrm{KS}(\mathbf{r},t)-v(\mathbf{r},t)$, is given by the Hartree (H) and the exchange-correlation (XC) potentials, which provide electrostatic and all further many-body effects, respectively.
According to the Runge-Gross theorem \cite{Runge84}, the XC potential is universal and depends on the full history of the electronic density and on the choice of initial states in the interacting and in the KS systems.
Incorporating these dependencies in calculations is a formidable task, see e.\,g.~\cite{Dobson94,Vignale95,Maitra01,Maitra02,Luo14,Fuks15}.
In practice, the XC potential is, therefore, almost exclusively approximated by adiabatic, i.\,e.\ time-local, density functionals, such as
the popular adiabatic local-density approximation.

\begin{figure}[b]
\begin{center}
\includegraphics[width=0.98\columnwidth]{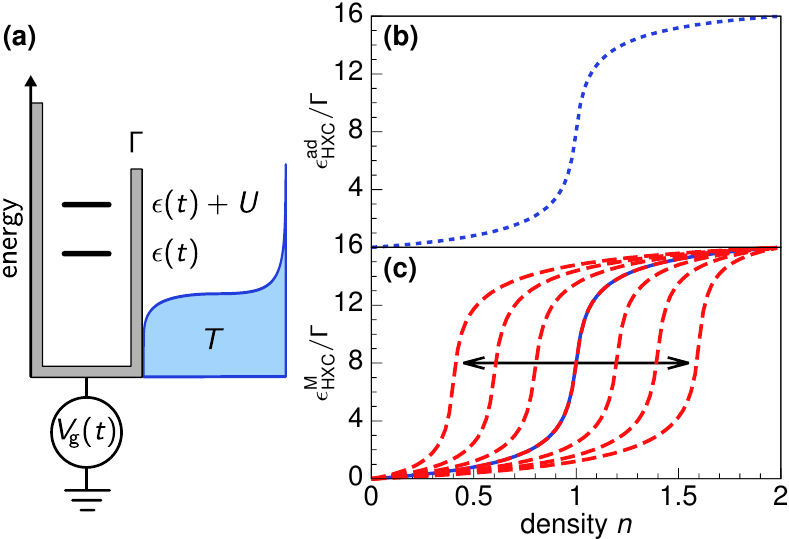}
\end{center}\vspace{-5mm}
	\caption{
	(a) Energy diagram of a single-level quantum dot with tunnel coupling, $\Gamma$, to a lead;
	(b) adiabatic HXC (XC plus Hartree) potential from \cite{Stefanucci11};
	(c) non-adiabatic HXC potential derived in Eq.~\eqref{eq_caseC_HXC}. Parameters are $U=16\Gamma$, $T=2\Gamma$
	and $\dot{n}/\Gamma = 1.2,0.8,\dots,-1.2$ (left to right lines). The black arrow indicates the dynamical step.
	}
	\label{fig_HXC}  
\end{figure}

In this Letter, we apply TDDFT to describe time-dependent charge transport in single-electron tunneling devices.
As a physical application, we analyze a single-electron source built by a quantum dot weakly tunnel-coupled to a nearby lead, see Fig.~\ref{fig_HXC} (a).
The emission of single charges into the lead is triggered by a time-periodic gate voltage.
By using insights from quantum-transport theory, we develop a non-adiabatic XC potential based on this system.
The developed XC potential displays features relevant for electron tunneling, in particular, in the presence of time-dependent driving. We then employ this XC potential for the time 
propagation of multiple quantum dots with weak tunnel coupling to a shared electron reservoir, see Fig.~\ref{fig_complexstruct} (a), to demonstrate its applicability to larger mesoscopic systems.
Both settings, the single-electron source and the multiple quantum dots, have been analyzed in recent experiments \cite{Feve07,Beckel14}.

The controlled emission of single electrons from a quantum dot into a solid-state device has been reported in Ref.~\cite{Feve07}, which, 
together with the advance of further single-electron sources \cite{Blumenthal07,Dubois13},
has sparked interest in the dynamics of mesoscopic systems at the single-particle level \cite{McNeil11,Hermelin11,Freulon15,Pekola13,Roche2013,Bocquillon13,Fletcher13}.
The basic physics of these systems is often well described by model Hamiltonians with Hubbard-like interaction terms, where the interaction parameters can be extracted from experiments.
However, even when relying on interacting model systems, it is a challenge to obtain the dynamics, see e.\,g.~\cite{Litinski17,Marguerite16}.
The TDDFT treatment of the dynamics of single-electron tunneling devices, which is outlined in this paper, thus mutually benefits the mesoscopic-transport and the TDDFT community:
First, it offers prospects of simulating time-resolved electron dynamics in interacting mesoscopic systems with TDDFT, 
which is numerically efficient, as it only requires a time propagation of non-interacting electrons.
Second, this approach points out non-adiabatic features of the universal XC potential, which are essential for an accurate TDDFT description.

The XC potential derived in this work has two key properties.
(1) For a stationary system, it reproduces the derivative discontinuity (DD)\mbox{---a} potential step at a~half-filled quantum-dot energy level \cite{Perdew82}.
The DD \cite{Perdew83,Mundt05,Lein05,Vieira09,Hellgren12} is relevant to reproduce Coulomb-blockade physics in a~non-interacting KS system \cite{Toher05,Kurth10,Evers11}.
(2) For the density on the quantum dot changing in time, e.\,g., due to a time-dependent gate voltage, the potential step shifts and appears at quantum-dot occupations which differ from the static case.
We find that this \emph{dynamical} step improves the TDDFT description by impeding electron tunneling in the KS system, see also Ref.~\cite{Elliott12a}.
Related dynamical steps have been reported for two-electron systems with long-range Coulomb interaction \cite{Lein05,Luo14,Hodgson16,Elliott12a} and for a 1D semiconductor \cite{Ramsden12}.
Importantly, we connect this step to electronic time scales by identifying charge relaxation rates of the single-electron source in the evolution of this step.

The quantum dot coupled to a lead, which acts as an electron reservoir, is illustrated in Fig.~\ref{fig_HXC}~(a).
The reservoir temperature is denoted by $T$ and electron-electron interaction in the reservoir is considered to be fully screened.
On the contrary, electrons occupying the quantum dot interact strongly due to their spatial confinement.
We take into account a single quantum-dot energy level, \mbox{$\epsilon(t)=-\alpha V_\mathrm{g}(t)$}, 
assuming a linear dependence on the applied gate voltage, $V_\mathrm{g}(t)$, with \mbox{$\alpha > 0$}. 
All energies are given with respect to the Fermi energy and $e, \hbar$, and $k_\mathrm{B}$ are set to one.
The system is modeled by the Anderson Hamiltonian
\begin{align}
\label{eq_hamiltonian}
\begin{aligned}[t]
 H = &\sum_\sigma \epsilon(t) d_\sigma^\dagger d_\sigma + U d_\uparrow^\dagger d_\uparrow d_\downarrow^\dagger d_\downarrow\\
     &+\sum_{k,\sigma} \epsilon_k c_{k\sigma}^\dagger c_{k\sigma} + \sum_{k,\sigma} \big( \gamma c_{k\sigma} d_{\sigma}^\dagger +  \mathrm{h.\,c.}\big),
\end{aligned}
\end{align}
where $d_\sigma$ $(d^{\dagger}_\sigma)$ denote the annihilation (creation) operators for quantum-dot states with spin index \mbox{$\sigma ={} \uparrow,\downarrow$}. 
The reservoir is described by a single energy band, $\epsilon_k$, whose annihilation (creation) operators for states with momentum $k$ and spin $\sigma$ are $c_{k\sigma}$ $(c^{\dagger}_{k\sigma})$.
We assume spin degeneracy for both the reservoir and the quantum dot and an energy-independent coupling, $\gamma$, between the two. 
The tunnel-coupling strength is defined as \mbox{$\Gamma = 2\pi |\gamma|^2 \nu_0$}, with $\nu_0$ being the density-of-states (DOS) at the Fermi energy.
The charging energy $U$ accounts for Coulomb repulsion on the quantum dot in the case of double occupation.

We first describe the time-dependent dynamics of this system by a Master equation.
In a second step, the obtained insights are applied to derive a non-adiabatic XC potential for the non-interacting KS system related to Eq.~\eqref{eq_hamiltonian}.
We consider the regime of weak reservoir-dot coupling, \mbox{$\Gamma \ll T$}, where the dynamics is Markovian whenever the reservoir memory time,
\mbox{$\tau_\mathrm{r} = T^{-1}$}, is smaller than time scales \cite{Splettstoesser06,Cavaliere09,Splettstoesser10,Riwar2016} introduced by the particular driving scheme.
In this regime, the dynamics of the single-electron source is well described by the sequential-tunneling picture \cite{Beenakker91},
which has the advantage of yielding an explicit analytic expression for the non-adiabatic XC potential derived below.
We note that our formalism can be readily extended to include higher-order tunneling corrections and also corrections with respect to non-Markovian dynamics,
by applying the real-time diagrammatic technique developed in Refs.~\cite{Koenig96a,Splettstoesser06,Cavaliere09,Splettstoesser10}.

We consider the reduced density matrix, $\rho_\mathrm{dot}(t)$, of the quantum dot, which is related to the full density matrix $\rho(t)$ of the reservoir-dot system by a partial trace over the reservoir degrees of freedom.
The reduced Hilbert space includes the many-particle states $\{\ket{0},\ket{\uparrow},\ket{\downarrow},\ket{2}\}$ denoting an empty, singly (with spin $\uparrow,\downarrow$) and doubly occupied dot.
The diagonal part of $\rho_\mathrm{dot}(t)$ defines the vector of occupation probabilities, \mbox{$\mathbf{P}(t) = [P_0(t),P_1(t),P_2(t)]^\mathrm{T}$}, where \mbox{$P_1(t)=P_\uparrow(t)+P_\downarrow(t)$}
captures both possible spin configurations for single occupation.
The electronic density on the quantum dot is \mbox{$n(t) = \mathbf{n}^\mathrm{T} \mathbf{P}(t)$}, with \mbox{$\mathbf{n}^\mathrm{T}=[0,1,2]$}.
We express the occupation vector in terms of the density, $n(t)$, and a further degree of freedom, $p(t)$, which is related to the parity of the quantum dot \cite{Schulenborg16}, by 
\begin{align}
 \label{eq_Pn}
 \mathbf{P}(t) = \left(\begin{matrix} 1-n(t)\\ n(t) \\0 \end{matrix}\right) + p(t) \left(\begin{matrix} 1\\-2\\1 \end{matrix}\right),
\end{align}
with \mbox{$n(t)/2 \geq p(t) \geq \max\big(0,n(t)-1\big)$}.
For our spin-degenerate system, the time evolution of the occupation vector decouples from the off-diagonal elements in $\rho_\mathrm{dot}(t)$ \cite{Koenig96a}, and is well approximated by the Master equation~\cite{commentInitialtime}
\begin{align}
  \label{eq_master_markov}
 \dot{\mathbf{P}}(t) &= \mathcal{W}(t)\mathbf{P}(t).
\end{align}
The kernel, $\mathcal{W}(t)$, is derived with parameters frozen at time $t$ and reads
\begin{equation}
\label{eq_caseA_W}
\frac{\mathcal{W}(t)}{\Gamma} = 
\left[
\begin{array}{ccc}
  -2f_\epsilon & f^-_\epsilon & 0 \\
  2 f_\epsilon & -f^-_\epsilon-f_{\epsilon+U} & 2 f^-_{\epsilon+U} \\
  0 & f_{\epsilon+U} & -2f^-_{\epsilon+U} 
\end{array}
 \right],
\end{equation}
with \mbox{$\epsilon=\epsilon(t)$}, the Fermi functions, \mbox{$f_\epsilon = 1/(1+e^{\epsilon/T})$} and \mbox{$f^-_\epsilon = 1-f_\epsilon$}. 
In the remainder of this paper, we exploit the well-known Eqs.~\eqref{eq_master_markov} and \eqref{eq_caseA_W} to develop a non-adiabatic XC potential for the non-interacting KS system related to Eq.~\eqref{eq_hamiltonian}.
In addition, we use them as a reference for the time evolution of the interacting system.

We now describe the dynamics of the single-electron source in TDDFT.
First, we make the usual assumption of non-interacting $v$-representability, i.\,e.,~the existence of a non-interacting KS system, which shares the same time-dependent electronic density with the interacting system 
\cite{MarquesMaitraNogueiraGrossRubio12,Ullrich11,Maitra16}.
The KS Hamiltonian is obtained from Eq.~\eqref{eq_hamiltonian} by setting $U$ to zero and by taking into account the H and XC potentials. We model the combined effect of both potentials
by an energy-level shift and hence $\epsilon(t)$ is replaced by $\epsilon(t)+\epsilon_{\mathrm{HXC}}[n](t)$ in Eq.~\eqref{eq_hamiltonian}.
Furthermore, we assume that $\epsilon_{\mathrm{HXC}}[n](t)$ solely depends on the electronic density on the quantum dot, $n(t)$, and its history \cite{commentXCbias}.

To derive a non-adiabatic approximation of the HXC potential, we express the dynamics of the KS system in terms of a Master equation.
This requires the additional assumption of Markovian dynamics to hold also in the KS system. Its time evolution is then given by
\begin{align}
 \label{eq_masterKS_markov}
 \dot{\mathbf{P}}_\mathrm{KS}(t) &= \mathcal{W}_\mathrm{KS}(t)\mathbf{P}_\mathrm{KS}(t),
 \end{align}
with the KS occupation vector, $\mathbf{P}_\mathrm{KS}(t)$, and the quantum-dot electronic density, \mbox{$n(t) = \mathbf{n}^\mathrm{T} \mathbf{P}_\mathrm{KS}(t)$}.
The KS kernel, $\mathcal{W}_\mathrm{KS}(t)$, is calculated from the expression in Eq.~\eqref{eq_caseA_W}
by setting \mbox{$U\rightarrow0$} and \mbox{$\epsilon(t)\rightarrow\epsilon(t)+\epsilon^{\mathrm{M}}_{\mathrm{HXC}}[n](t)$}. 
The superscript `M' reminds that the Markov approximation is applied in Eqs.~\eqref{eq_master_markov} and \eqref{eq_masterKS_markov}.

Based on the Master equations \eqref{eq_master_markov} and \eqref{eq_masterKS_markov},
we now derive a non-adiabatic approximation of the HXC potential.
The key observation is that the position of the quantum dot energy level is uniquely fixed by the density and its first time-derivative.
In the interacting system, the latter is written, with Eq.~\eqref{eq_master_markov}, as \mbox{$\dot{n}(t) = \mathbf{n}^\mathrm{T} \mathcal{W}(t) \mathbf{P}(t)$}.
Inserting the representation of the occupation vector \eqref{eq_Pn} into this equation, only the contribution stemming from the first term on the r.\,h.\,s.~of Eq.~\eqref{eq_Pn} remains \cite{Schulenborg16}.
This means that the time-dependent energy level of the quantum dot, $\epsilon(t)$, only depends on $n(t)$ and $\dot{n}(t)$. Explicitly, we write $\epsilon(t)$ as a function, $g$, with
\begin{align}
  \label{eq_g}
  \epsilon(t) &= g\big(n(t),\dot{n}(t),\Gamma,T,U\big),
\end{align}
where $\Gamma$, $T$ and $U$ are constants.
The same inversion in the non-interacting KS system leads to
\begin{align}
  \label{eq_gKS}
  \epsilon(t)+\epsilon_\mathrm{HXC}^{\mathrm{M}}[n](t) &= g\big(n(t),\dot{n}(t),\Gamma,T,0\big),
\end{align}
where we have used the fact that the KS and the interacting system have the same density, $n(t)$. Solving Eqs.~\eqref{eq_g} and \eqref{eq_gKS} for $\epsilon_{\mathrm{HXC}}^{\mathrm{M}}$, we obtain the first key result,
\begin{subequations}
\label{eq_caseC_HXC}
\begin{align}
\epsilon_{\mathrm{HXC}}^{\mathrm{M}}\big(n(t),\dot{n}(t)\big)(t) &= T \log \Big\{C\big(n(t),\dot{n}(t)\big)\Big\},
\end{align}
with
\begin{align}
\lefteqn{\frac{1}{C(n,\dot{n})} = \frac{\dot{n} + e^{U/T}\big(\dot{n}+2\Gamma (n-1)\big)}{2 e^{U/T}\big(\dot{n}+\Gamma(n-2)\big)}}\hspace{0.3cm}\\
\nonumber
&\times&\bigg\{1 - \bigg(1
-\frac{4e^{U/T}\left((\dot{n}+\Gamma n)^2-2\Gamma (\dot{n}+\Gamma n)\right)}{\big\{\dot{n} +e^{U/T}\big(\dot{n}+2\Gamma (n-1)\big)\big\}^2}\bigg)^{\!\!\frac{1}{2}}\bigg\}.
\end{align}
\end{subequations}
Equation \eqref{eq_caseC_HXC} defines an HXC potential, which is non-adiabatic due to its dependence on the first time-derivative of the density.
We remark that a general functional dependence is denoted by $[n]$, but, in our specific approximation, the HXC potential becomes a function of $n(t)$ and $\dot{n}(t)$.
Importantly, for vanishing $\dot{n}(t)$, the expression in Eq.~\eqref{eq_caseC_HXC}
equals the adiabatic HXC potential derived in Ref.~\cite{Stefanucci11}, \mbox{$\epsilon_{\mathrm{HXC}}^{\mathrm{ad}}\big(n(t)\big)(t) = \epsilon_{\mathrm{HXC}}^{\mathrm{M}}\big(n(t),0\big)(t)$}, 
which we use for comparison.

To outline the non-adiabatic property of the derived HXC potential in Eq.~\eqref{eq_caseC_HXC}, we first compare the form of both the adiabatic and the non-adiabatic HXC potentials in Fig.~\ref{fig_HXC} (b)-(c).
As visible in (b), the adiabatic potential shows a step, previously denoted as DD, which is centered at the electron-hole symmetric point, \mbox{$n=1$}, and smeared out by temperature \cite{Stefanucci11,Evers11,Xianlong12}.
This feature is strikingly modified in the non-adiabatic HXC potential, shown in Fig.~\ref{fig_HXC} (c). Here, a non-zero time derivative of the density shifts the position of the step to different density values, 
giving rise to a dynamical step. 
This dynamical step clearly emerges from the DD of the stationary system.
The position of the dynamical step as a function of the time-derivative of the density reads as
\begin{align}
\label{eq_DDD}
n = 1-\dot{n} \left(\tau_\mathrm{c}^{U=0}  - \tau_\mathrm{c}^{U\neq 0} \right),
\end{align}
where \mbox{$\tau_\mathrm{c}^{U=0} = \Gamma^{-1}$} denotes the time scale of charge relaxation in the non-interacting system, while \mbox{$\tau_\mathrm{c}^{U\neq0} = \Gamma^{-1}_\mathrm{c}(-U/2)$} denotes 
the respective time scale in the interacting system, evaluated at electron-hole symmetry. 
The charge relaxation rate of the interacting system is given by \mbox{$\Gamma_\mathrm{c}(\epsilon)/\Gamma = 1+f_\epsilon-f_{\epsilon+U}$} \cite{Cavaliere09,Splettstoesser10}.
This link of the dynamical step to \emph{physical} time scales of electron dynamics is the second key result of this work. 
Furthermore, Eq.~\eqref{eq_DDD} shows that only if the second term on the r.\,h.\,s.~is small during the full time propagation, \mbox{$|\dot{n} |\ll |\tau_\mathrm{c}^{U=0} - \tau_\mathrm{c}^{U\neq 0}|^{-1}$}, 
the dynamical step is always close to \mbox{$n=1$} and the adiabatic HXC potential, $\epsilon_\mathrm{HXC}^{\mathrm{ad}}$, becomes a sufficient approximation.

We now investigate the performance of the non-adiabatic HXC potential developed here in TDDFT simulations of the single-electron source.
Therefore, we assume a finite but sufficiently large number of states for the reservoir and numerically propagate the KS density matrix of the combined 
reservoir-dot system in time until periodicity has been established \cite{supp}.
Our ensemble TDDFT calculations begin with an equilibrium KS density matrix with temperature $T>0$.
Since the KS system is non-interacting, this involves the time propagation of single-particle wave functions using a continuously updated KS Hamiltonian.
We emphasize that the HXC potential is the only approximation in a TDDFT calculation, while no further approximations, e.\,g.,~with respect to
weak tunnel coupling or to Markovian dynamics, are made for the time propagation.
Our results are compared to the ones obtained by applying the adiabatic HXC potential as well as Eq.~\eqref{eq_master_markov}.

\begin{figure}[t]
\begin{center}
\includegraphics[width=0.98\columnwidth]{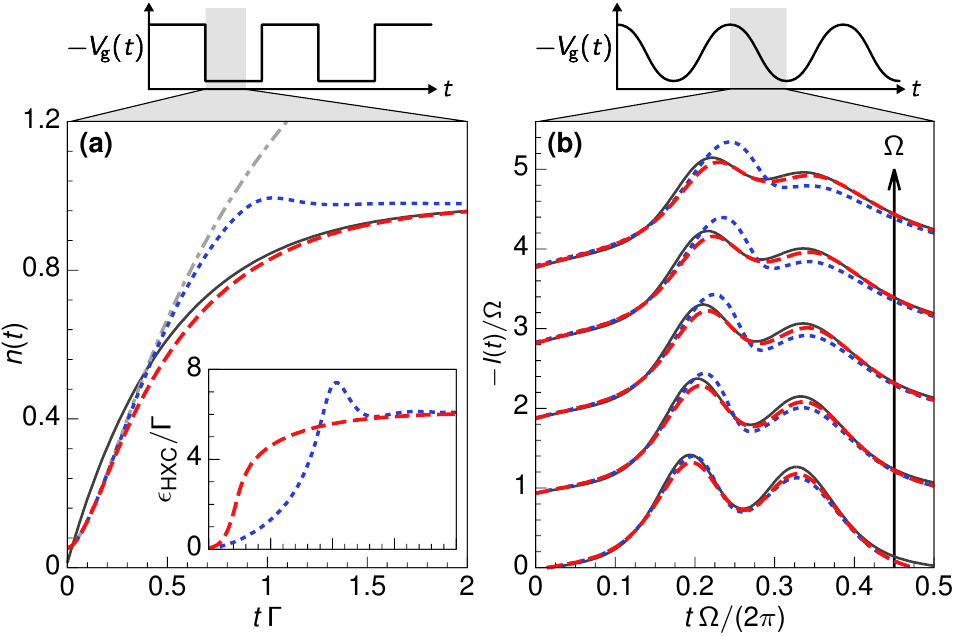}
\end{center}\vspace{-4mm}
	\caption{	
 	Comparison of TDDFT results obtained
 	using $\epsilon_{\mathrm{HXC}}^{\mathrm{ad}}$ (dotted lines), $\epsilon_{\mathrm{HXC}}^{\mathrm{M}}$ (dashed lines) and no HXC potential (dashed-dotted lines)
	with analytic results from Eq.~\eqref{eq_master_markov} (solid lines);
	(a) densities of the quantum dot subject to a square pulse (inset:~HXC potentials) with
	parameters \mbox{$U=16\Gamma$}, \mbox{$T=2\Gamma$}, \mbox{$\epsilon(t<0)=10\Gamma$}, \mbox{$\epsilon(t>0)=-6\Gamma$};
	(b) currents in the case of a harmonic gate-voltage drive plotted for driving frequencies in the range $[0.1\Gamma,0.5\Gamma]$ with \mbox{$\Delta \Omega = 0.1\Gamma$}
	(note that for visibility lines are shifted by 1). Further parameters are \mbox{$U=12\Gamma$}, \mbox{$T=3\Gamma$}, \mbox{$\epsilon(t)=-6\Gamma+18\Gamma\cos(\Omega t)$}.
	}
	\label{fig_quenchsin} 
\end{figure}

\begin{figure}[t]
\begin{center}
\includegraphics[width=0.98\columnwidth]{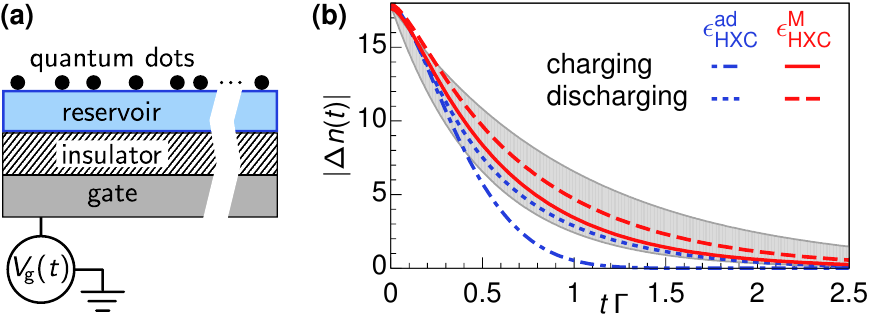}
\end{center}
	\caption{	
 	(a) Setup of 70 quantum dots tunnel-coupled to the same electron reservoir;
 	(b) absolute density change summed over all quantum dots, \mbox{$|\Delta n(t)| = \sum_{l=1}^{70} \big|n_l(t)-n_l(t_1)\big|$} with \mbox{$t_1 \gg t$}, 
 	for a sudden rise (charging) or drop (discharging) of the gate voltage at \mbox{$t=0$}, 
 	calculated using either $\epsilon^\mathrm{M}_\mathrm{HXC}$ or $\epsilon^\mathrm{ad}_\mathrm{HXC}$ for all dots.
	See text for gray area and Ref.~\cite{supp} for parameters.
	}
	\label{fig_complexstruct} 
\end{figure}

First, we analyze the time-dependent charge flow between the quantum dot and the reservoir induced by a step-pulse gate-voltage driving, see Fig.~\ref{fig_quenchsin}~(a).
For weak reservoir-dot coupling, as considered here, the time-dependent density, $n(t)$, is well described by an exponential decay towards its new equilibrium value after each gate-voltage step.
The characteristic decay rate is given by $\Gamma_\mathrm{c}(\epsilon)$ \cite{Cavaliere09,Splettstoesser10}, where $\epsilon$ is the new position of the energy level.
This is shown by the solid line in Fig.~\ref{fig_quenchsin} (a) and compared to the calculated TDDFT densities.
Interestingly, the density related to $\epsilon_{\mathrm{HXC}}^{\mathrm{ad}}$ 
first follows the dashed-dotted line, which shows the exponential decay for a system without interaction (\mbox{$U=0$}) obtained from a TDDFT calculation with \mbox{$\epsilon_{\mathrm{HXC}}=0$}. 
Thus, we conclude that the adiabatic HXC potential leads to charge relaxation with the characteristic decay rate of a non-interacting system.
The density eventually overestimates its equilibrium value and strongly decaying oscillations occur, which are visible in the inset (dotted line) of Fig.~\ref{fig_quenchsin}~(a), 
showing the time evolution of the adiabatic HXC potential.
On the contrary, the TDDFT density in Fig.~\ref{fig_quenchsin}~(a), which corresponds to the non-adiabatic HXC potential $\epsilon_{\mathrm{HXC}}^{\mathrm{M}}$ (dashed)
clearly features the expected exponential behavior with the decay rate of the interacting system.
The time evolution of $\epsilon_{\mathrm{HXC}}^{\mathrm{M}}$ in the inset of Fig.~\ref{fig_quenchsin}~(a) also indicates that the density increases monotonically towards the new equilibrium value without any oscillating behavior.
It is remarkable that the dependence on $\dot{n}(t)$ in the HXC potential is already sufficient to generate the charge-relaxation behavior of an interacting system in a non-interacting KS system.
We also checked that at the time at which the two HXC potentials begin to differ, \mbox{$t \Gamma \gtrsim 1/4$}, the respective TDDFT densities in Fig.~\ref{fig_quenchsin}~(a) evolve with different decay rates.
Finally, we note that minor differences between the TDDFT and the Master-equation data, which are visible in the beginning of the time evolution
are related to non-Markovian dynamics, which is fully neglected in Eq.~\eqref{eq_master_markov} but partially included in a TDDFT calculation.

As a second application, we analyze a large-amplitude harmonic driving of the gate voltage, see Fig.~\ref{fig_quenchsin}~(b), presenting the resulting time-resolved currents, \mbox{$I(t)=-\dot{n}(t)$}.
The two peaks correspond to the first and the second electron entering the initially empty quantum dot during a half period of the drive.
The frequency increases linearly from bottom to top lines and all further parameters are chosen to illustrate the transition from the adiabatic to the non-adiabatic regime.
The solid lines represent the result of Eq.~\eqref{eq_master_markov} and serve as a reference point, allowing
for a comparison with the TDDFT currents related to the adiabatic HXC potential, which reveals the breakdown of $\epsilon_{\mathrm{HXC}}^{\mathrm{ad}}$ for driving beyond the adiabatic regime. 
For larger frequencies, the left peak in the charge current is increasingly overestimated, while the right peak is underestimated.
The poorly reproduced charge relaxation rate causes electrons in the KS system of the adiabatic HXC potential to tunnel too quickly.
In contrast, the non-adiabatic potential leads to a good agreement between the TDDFT currents and the result of Eq.~\eqref{eq_master_markov} for all displayed driving frequencies.

Third, we turn to a larger structure and analyze the experiment reported in Ref.~\cite{Beckel14}, where a self-assembled layer of quantum dots is located on top of a two-dimensional electron gas. 
We model this setup by weakly coupling 70 quantum dots to a shared electron reservoir as shown in Fig.~\ref{fig_complexstruct} (a).
To include parameter variations of a real setup, we consider Gaussian-distributed energy levels ($\epsilon_l$), reservoir-dot couplings ($\gamma_l$) and interaction strengths ($U_l$) for the quantum dots with site index $l$.
The spatial distances between adjacent dots are modeled by multiplying $\gamma_l$ with an energy-dependent phase factor \cite{Alexander64,supp}.
In the experiment, these distances are larger than the reservoir coherence length, which justifies the application of the 
HXC potentials $\epsilon_{\mathrm{HXC}}^{\mathrm{M/ad}}$ for each dot separately.
Figure~\ref{fig_complexstruct}~(b) presents TDDFT calculations for a step-pulse gate-voltage driving, leading to the periodic charging/discharging of a fraction of quantum dots.
While for both HXC potentials we observe the overall signature~\cite{Beckel14} of interaction, namely that
the relaxation dynamics after positive/negative gate-voltage steps differ,
the explicit evolution is \emph{physical} only for~$\epsilon^\mathrm{M}_\mathrm{HXC}$.
For times exceeding the reservoir memory time, $\tau_\mathrm{r}$,
the density changes in Fig.~\ref{fig_complexstruct}~(b) have to lie in the gray region bounded by $|\Delta n(0)|e^{-\Gamma t}$ and $|\Delta n(0)|e^{-2\Gamma t}$, 
since \mbox{$\Gamma\leq\Gamma_\text{c}(\epsilon)\leq2\Gamma$} for a single dot.
The charging process calculated with $\epsilon^\mathrm{ad}_\mathrm{HXC}$ violates this constraint,
because each dot which is supposed to become singly occupied evolves towards double occupation initially, see also Fig.~\ref{fig_quenchsin}~(a).
We emphasize that our TDDFT treatment can include the geometry of interest and therefore allows for future analyses beyond the statistical method applied in Ref.~\cite{Beckel14}, e.\,g.,~with reduced quantum-dot distances.
We also note that the numerical costs of time evolving the setups of Figs.~\ref{fig_quenchsin} and~\ref{fig_complexstruct} are comparable.

To summarize, we applied TDDFT to time evolve a single-electron source in the presence of Coulomb interaction.
For this purpose, we developed a non-adiabatic HXC potential, with the key feature of a dynamical step, which shifts if the electronic density changes in time.
We provided an explicit link between density values where the dynamical step occurs and charge-relaxation time scales evaluated at the electron-hole symmetric point.
Since the exact HXC potential is a universal quantity, our results have relevance beyond the single-electron source studied here. 
We already demonstrated the applicability of our approach to a structure containing multiple quantum dots.
Related studies of other complex mesoscopic systems, e.\,g.,~including several leads, are in reach.


We thank N.~Maitra, M.~Misiorny, J.~Odavi\'c and J.~Schulenborg for valuable comments on this manuscript.
We further enjoyed discussions with J.~Fuks, P.~Hyldgaard, L.~Lacombe and E.~Schr\"oder.
Support from the Deutsche Forschungsgemeinschaft via RTG1995 (ND, JS) and an Emmy-Noether grant (NH), as well as 
from the Knut and Alice Wallenberg foundation and the Swedish VR (JS), is gratefully acknowledged.



\begin{thebibliography}{53}
\bibitem [{com({\natexlab{a}})}]{commentInitialtime}%
  \BibitemOpen
  \bibinfo {note} {The initial time has been sent to minus infinity and any
  information of the initial condition has thereby decayed.}\BibitemShut
  {Stop}%
\bibitem [{com({\natexlab{b}})}]{commentXCbias}%
  \BibitemOpen
  \bibinfo {note} {We note that in a two-lead setup, not considered here, one
  would need to account for XC corrections to an applied bias voltage
  \cite{Kurth13,Stefanucci15,Liu15,Stefanucci17}.}\BibitemShut {Stop}%
\end{thebibliography}

%

\onecolumngrid
\clearpage


\section*{Supplementary material}

# Supplemental Material for 'Non-Adiabatic Dynamics in Single-Electron Tunneling Devices with Time-Dependent Density Functional Theory'

Niklas Dittmann[(1,2,3)], Janine Splettstoesser[(1)], and Nicole Helbig[(3)]

(1) Department of Microtechnology and Nanoscience (MC2),
Chalmers University of Technology,
SE-41298 Göteborg, Sweden

(2) Institute for Theory of Statistical Physics,
RWTH Aachen, 52056 Aachen, Germany

(3) Peter-Grünberg Institut and Institute for Advanced Simulation,
Forschungszentrum Jülich,
52425 Jülich, Germany

We outline our implementation of time-dependent density functional theory (TDDFT) for $N_\mathrm{dot}$ interacting single-level quantum dots coupled to an electron reservoir. The Hamiltonian reads

$$H = \sum_{l,\sigma} \epsilon_l(t) d^\dagger_{l\sigma} d_{l\sigma} + \sum_l U_l d^\dagger_{l\uparrow} d_{l\uparrow} d^\dagger_{l\downarrow} d_{l\downarrow} + \sum_{k,\sigma} \epsilon_k c^\dagger_{k\sigma} c_{k\sigma} + \left( \sum_{k,l,\sigma} \gamma_{kl} c_{k\sigma} d^\dagger_{l\sigma} + \mathrm{h.\,c.} \right), \quad (1)$$

with the quantum-dot index $l = 1, \ldots, N_\mathrm{dot}$ and where otherwise the notation of the main paper has been adopted (see Eq. (1) in the latter). Furthermore, for $N_\mathrm{dot} > 1$, the real-space positions $\{r_l\}$ of the quantum dots are modeled by an energy-dependent phase of the reservoir-dot couplings, $\gamma_{kl} = \gamma_l e^{ikr_l}$, see e. g. Ref. [1]. We note that this modeling allows to analyze various spatial geometries of quantum dots coupled to a shared electron reservoir. We define the Hamiltonian of the related non-interacting Kohn-Sham (KS) system [2, 3] as

$$H_\mathrm{KS} = \sum_{l,\sigma} \left( \epsilon_l(t) + \epsilon_{\mathrm{HXC},l}[n](t) \right) d^\dagger_{l\sigma} d_{l\sigma} + \sum_{k,\sigma} \epsilon_k c^\dagger_{k\sigma} c_{k\sigma} + \left( \sum_{k,l,\sigma} \gamma_{kl} c_{k\sigma} d^\dagger_{l\sigma} + \mathrm{h.\,c.} \right). \quad (2)$$

The total electronic density on the quantum dots is calculated, as $n(t) = \sum_{l,\sigma} \left\langle d^\dagger_{l\sigma} d_{l\sigma} \right\rangle(t)$, by time propagating KS single-particle orbitals. We apply a time-propagation scheme based on an open-boundary Schrödinger equation (OBSE) [4]. To this end, we partition the KS Hamiltonian, expressed in the single-particle basis given by the KS orbitals of quantum dots and reservoir, as

$$H_\mathrm{KS} = \begin{pmatrix} H_\mathrm{dd} & H_\mathrm{dr} \\ H_\mathrm{rd} & H_\mathrm{rr} \end{pmatrix}, \quad (3)$$

where $H_\mathrm{dd}$ ($H_\mathrm{rr}$) acts on the quantum-dots (reservoir) degrees of freedom, while $H_\mathrm{dr}$ and $H_\mathrm{rd}$ describe tunneling between the quantum dots and the reservoir. Since the spin sector factorizes for

the KS Hamiltonian in Eq. (2), we restrict to a single spin species in the following discussion. As outlined in Ref. 4, the time-dependent Schrödinger equation for a KS single-particle wave function, $\psi(t)$, projected onto the quantum-dots region, $\psi_{\mathrm{d}}(t)$, can be written as

$$i\dot{\psi}_{\mathrm{d}}(t) = H_{\mathrm{dd}}(t)\psi_{\mathrm{d}}(t) + \int_{t_0}^{t} dt' \Sigma(t,t')\psi_{\mathrm{d}}(t') + iH_{\mathrm{dr}}g_{\mathrm{r}}(t,t_0)\psi_{\mathrm{r}}(t_0), \qquad (4)$$

with $\Sigma(t,t') = H_{\mathrm{dr}}g_{\mathrm{r}}(t,t')H_{\mathrm{rd}}$ and the reservoir Greens function $g_{\mathrm{r}}(t,t')$. The two additional terms on the r. h. s. of Eq. (4), respectively the memory and the source term, are due to the presence of the reservoir, and $\psi_{\mathrm{r}}(t_0)$ is the projection of the KS single-particle wave function onto the reservoir region, evaluated at the initial time $t_0$. The reservoir Greens function is defined as the solution of the equation $(i\partial_t - H_{\mathrm{rr}})g_{\mathrm{r}}(t,t') = \delta(t-t')$, with initial value $g_{\mathrm{r}}(t^+,t) = -i$, which is

$$g_{\mathrm{r}}(t,t') = \frac{1}{2\pi}\int d\omega e^{-i\omega(t-t')}(\omega - H_{\mathrm{rr}} + i\eta)^{-1}, \qquad (5)$$

with $\eta \to 0$. Inserting the partitioned Hamiltonian and the reservoir Greens function into Eq. (4), we obtain, for the $l$'th entry,

$$i\dot{\psi}_{\mathrm{d},l}(t) = \left[\epsilon_l(t) + \epsilon_{\mathrm{HXC},l}[n](t)\right]\psi_{\mathrm{d},l}(t) - i\int_{t_0}^{t} dt' \sum_{l',k} \gamma_l \gamma_{l'}^* e^{ik(r_l - r_{l'}) + i\epsilon_k(t'-t)}\psi_{\mathrm{d},l'}(t') \qquad (6)$$
$$+ \sum_{k} \gamma_{kl} e^{-i\epsilon_k(t-t_0)}\psi_{\mathrm{r},k}(t_0),$$

where $\psi_{\mathrm{r},k}(t_0)$ is the $k$'th component of the vector $\psi_{\mathrm{r}}(t_0)$. Since in the main paper, we assume a reservoir with a one-dimensional flat energy band, we rewrite the product $k(r_l - r_{l'})$ in Eq. (6) as $\mathrm{sgn}(k)\epsilon_k(\bar{t}_l - \bar{t}_{l'})$, with $\bar{t}_l = \frac{|r_l|}{v_{\mathrm{F}}}$, the Fermi velocity $v_{\mathrm{F}}$ and with sgn being the sign function. Moreover, we write the first sum over $k$ as an integral over a constant density of states, which leads to a delta distribution. We derive the final expression,

$$i\dot{\psi}_{\mathrm{d},l}(t) = \left[\epsilon_l(t) + \epsilon_{\mathrm{HXC},l}[n](t) - \frac{i\Gamma_l}{2}\right]\psi_{\mathrm{d},l}(t) + \sum_{k}\gamma_l e^{i\mathrm{sgn}(k)\epsilon_k \bar{t}_l - i\epsilon_k(t-t_0)}\psi_{\mathrm{r},k}(t_0) \qquad (7)$$
$$- \frac{i}{2}\sum_{l\neq l'}\Gamma_{ll'}\psi_{\mathrm{d},l'}(t - |\bar{t}_l - \bar{t}_{l'}|)\theta(t - t_0 - |\bar{t}_l - \bar{t}_{l'}|),$$

with $\Gamma_{ll'} = 2\pi\nu_0 \gamma_l \gamma_{l'}^*$, the tunnel-coupling strengths $\Gamma_l = \Gamma_{ll}$, the density of states at the Fermi energy $\nu_0$ and with $\theta$ denoting the Heaviside function. For $N_{\mathrm{dot}} = 1$ we furthermore define $\Gamma = \Gamma_1$. Eq. (7) has been applied in our TDDFT calculations. Notably, the second line in Eq. (7) only contributes if the propagation time is larger than $|\bar{t}_l - \bar{t}_{l'}|$ for any $l' \neq l$.

As initial condition at time $t_0$, we consider the equilibrium density matrix of the uncoupled KS system ($\gamma_l = 0$). The initial occupation of the reservoir orbital related to $\epsilon_k$ is, therefore, given by

$f(\epsilon_k)$ in the spin-less system, and $2f(\epsilon_k)$ in the spin-full system, where $f(\epsilon) = 1/(1+e^{\epsilon/T})$ denotes the Fermi function with temperature $T$ and $k_\mathrm{B} = 1$. The initial occupations of the quantum dots have been chosen arbitrarily in our calculations, because the TDDFT densities presented in the main paper are all shown for $t \gg t_0$, where any current due to initial quantum-dot occupations has decayed. The density on the quantum dot $l$ reads $n_l(t) = \sum_{\{\psi\}} p_\psi |\psi_{\mathrm{d},l}(t)|^2$, where the sum goes over all KS orbitals and $p_\psi$ is the initial occupation of each orbital. In practice, we use a discrete number of KS orbitals by considering reservoir energy levels, $\epsilon_k$, equally distributed over an energy interval, $I_\mathrm{R}$, in accordance with the flat-band approximation. The KS orbitals are numerically propagated in time by applying Eq. (7). After each time step, the HXC potential, approximated by either $\epsilon_\mathrm{HXC}^\mathrm{M}$ or $\epsilon_\mathrm{HXC}^\mathrm{ad}$ for each quantum dot separately, is adjusted according to the updated densities. The TDDFT densities shown in Fig. 2 (left) in the main paper have been obtained by propagating 1600 reservoir orbitals, distributed in $I_\mathrm{R} = [-900\Gamma, 900\Gamma]$, with a time step $\Delta t = 10^{-5}/\Gamma$. The TDDFT currents in Fig. 2 (right) have been calculated with 1000 reservoir orbitals (lowest drive frequency with 1300 orbitals), $I_\mathrm{R} = [-100\Gamma, 100\Gamma]$ and $\Delta t = 3 \cdot 10^{-5}/\Gamma$. For Fig. 3, we applied 1600 orbitals, $I_\mathrm{R} = [-200\Gamma, 200\Gamma]$ and $\Delta t = 3 \cdot 10^{-5}/\Gamma$. Moreover, to calculate this figure we used $N_\mathrm{dot} = 70$ quantum dots with Gaussian-distributed energy levels $\epsilon_l$ ($\mu = 0.7\Gamma, \sigma = 0.5\Gamma$), interaction strengths $U_l$ ($\mu = 14\Gamma, \sigma = 0.5\Gamma$) and reservoir-dot couplings $\gamma_l$ ($\mu = 1\Gamma, \sigma = 0.1\Gamma$), where $\Gamma$ defines an energy scale. The applied square-pulse gate voltage abruptly shifts all quantum-dot energy levels by the amount $-3\Gamma$ (charging) and back (discharging). We furthermore considered the distances between neighboring dots to be sufficiently large to drop the term in the second line of Eq. (7) in our time evolution. Notably, we expect the importance of this term to be reduced by processes which lead to decoherence in the reservoir, e. g. electron-electron or electron-phonon scattering, which are not included in our calculations.

---



\end{document}